\documentclass[conference]{IEEEtran}
\ifCLASSINFOpdf
   \usepackage[pdftex]{graphicx}
  \graphicspath{{../pdf/}{../jpeg/}}
\else
\fi
\usepackage{amsmath}

\usepackage{alltt                                    
          , multirow
          , booktabs
          , listings
          , graphicx
          ,float
	,cite
          ,verbatim
         ,mathtools
	,url
	,amsmath
}
\usepackage[T1]{fontenc}
\usepackage[table]{xcolor}
\usepackage[numbers]{natbib}     
\usepackage{syntax}
\usepackage{algorithmic, algorithm}
\usepackage{enumitem}
\usepackage{threeparttable}
\usepackage{natbib}
\usepackage{hhline}
\usepackage{balance}
\usepackage{framed}

\usepackage{expl3}
\ExplSyntaxOn
\newcommand\latinabbrev[1]{
  \peek_meaning:NTF . {
    #1\@}%
  { \peek_catcode:NTF a {
      #1., \@ }%
    {#1., \@}}}
\ExplSyntaxOff

\newcommand{\CASE}[1]{\STATE \textbf{case} #1\textbf{:} \begin{ALC@g}}
\newcommand{\ENDCASE}{\end{ALC@g}}

\newcommand{\DEFAULT}{\STATE \textbf{default:} \begin{ALC@g}}
\newcommand{\ENDDEFAULT}{\end{ALC@g}}
\newcommand{\DEFAULTLINE}[1]{\STATE \textbf{default:} }
\let\footnotesize\scriptsize

\newcommand{\cfbox}[2]{%
    \colorlet{currentcolor}{.}%
    {\color{#1}%
    \fbox{\color{currentcolor}#2}}%
}

\newsavebox{\supbox}
\newcommand{\bsup}{\begin{lrbox}{\supbox}$\tt\scriptstyle}
\newcommand{\esup}{$\end{lrbox}{}^{\usebox{\supbox}}}
\def\eg{\latinabbrev{e.g}}
\def\ie{\latinabbrev{i.e}}

\definecolor{lightpurple}{rgb}{0.8,0.8,1}
\definecolor{codebg}{RGB}{255,255,255}
\definecolor{commentcolor}{RGB}{11,140,11}
\definecolor{lightgray}{RGB}{220,220,220}
\newcommand{\hl}{\makebox[0pt][l]{\color{lightgray}\rule[-2.5pt]{0.90\linewidth}{9pt}}}
\colorlet{shadecolor}{blue!5}

\begin{document}

%
\title{On the Use of Context in Recommending Exception Handling Code Examples}


\author{\IEEEauthorblockN{Mohammad Masudur Rahman~~~~~ Chanchal K. Roy }
\IEEEauthorblockA{Department of Computer Science, University of Saskatchewan, Canada\\
\{masud.rahman, chanchal.roy\}@usask.ca}
}



%


\maketitle


\begin{abstract}
Studies show that software developers often either misuse exception handling features or use them inefficiently, and such a practice may lead an undergoing software project to a fragile, insecure and non-robust application system. In this paper, we propose a context-aware code recommendation approach that recommends exception handling code examples from a number of popular open source code repositories hosted at GitHub. It collects the code examples exploiting \emph{GitHub code search API}, and 
then analyzes, filters and ranks them against the code under development in the IDE by leveraging not only
the \emph{structural} (\ie\ graph-based) and \emph{lexical} features but also the \emph{heuristic quality measures} of exception handlers in the examples. 
Experiments with 4,400 code examples and 65 exception handling scenarios as well as comparisons with four existing approaches show that the proposed approach is highly promising. 



\end{abstract}


%

\IEEEpeerreviewmaketitle

\section{Introduction}
Exception handling is one of the most important tasks that software developers undertake during software development and maintenance.
However, studies show that developers either use the exception-handling features ineffectively \cite{fieldstudy} or misuse them in the real life software development \cite{shah}. \citet{fieldstudy} conduct a study with 32 applications from Java and .NET frameworks, and report that about 40\%-70\% exception handling actions are overly simplified or ineffective. The actions either log error messages and print stack traces or simply do nothing. According to their findings, developing effective exception handlers is a daunting task.
One way to benefit both the developers' productivity and the quality of the exception handlers is to recommend readily available and relevant exception handling code examples to the developers within the scope of their working environment (\eg\ IDE), which can be leveraged in handling exceptions by them.

A number of existing studies on exception handling attempt to support the developers through useful insights from static analysis of the exception control flows and handling structures \cite{robi, goodenough, chang} or comparative field studies \cite{fieldstudy, garcia}, visualization \cite{shah}, and recommendation of code examples \cite{heuristics}. \citet{heuristics} propose an approach to recommend exception handling code examples exploiting the structural facts of the code under development and the candidate examples. 
Although the approach performs considerably well on their carefully constructed dataset, it suffers from several limitations. 
First, the approach considers only the usage of certain API classes and API methods, and captures neither static relationships (\ie\ method belongs to which class) nor the dependencies among different API objects used in the code. These static or dependency relationships can be considered as an important structural component of a code example. Second, the constructed dataset is static and cannot be easily updated. It also requires significant amount of manual preprocessing to be useful in the recommendation for exception handling.
 
 In this paper, we propose a \emph{context-aware} recommendation approach for exception handling code examples, which leverages not only both the \emph{structural} and \emph{lexical} features but also the \emph{quality of the exception handlers} in the code examples. The approach exploits the \emph{GitHub Code Search API} \cite{gitapi}, and collects about 60-70 code examples from GitHub code repositories against a search query representing the context code (\ie\ code under development) in the IDE and the exception a developer attempts to handle. It then analyzes, filters and ranks the examples based on their relevance against the context code and the quality of their exception handlers.

The proposed approach also overcomes certain limitations of the existing techniques. 
First, it adopts a \emph{graph-based technique} for structural relevance estimation, where the approach identifies all the API objects along with their static relationships and data dependencies in the code to develop an \emph{API usage graph}.
We believe that two code fragments having similar usage graphs (\ie\ similar set of API objects with similar static or dependency relationships) are likely to accomplish similar programming tasks. 
The usage graph captures  more useful and more in-depth structural features of the code compared to existing structural heuristics \cite{heuristics, strathcona}. We thus exploit the usage graph matching idea for structural relevance estimation (\ie\ novelty of our approach), and it helps to overcome the limitations of the heuristic-based techniques.
Second, it applies a state-of-the-art lexical feature-based code cloning technique \cite{croy} in order to determine the \emph{lexical similarity} between context code in the IDE and the candidate examples, which was completely ignored by the existing studies. The idea is to recommend the code examples which are not only structurally relevant but also lexically similar (\ie\ easy to work with) to the context code. Third, the approach integrates one of the largest and the most popular online code bases, \emph{GitHub}, into the IDE, which can provide readily available exception handling code examples from the top ranked repositories. This integration makes the corpus for recommendation dynamic, constantly evolving, and synchronized with a number of mature and popular projects.


We conduct experiments on the proposed approach using 4,400 GitHub code examples and 65 exception handling scenarios (\ie\ each scenario consists of an exception and a code segment). The exceptions and associated context code are collected from different online sources such as StackOverflow Q \& A site and Pastebin \cite{pb}. First, we perform an extensive search into GitHub code repositories using the code search feature, and develop an \emph{oracle} by collecting the most relevant exception handling code examples for each case (\ie\ scenario). We then use the oracle in order to evaluate the proposed approach, where our approach recommends relevant code examples with a maximum of 41.92\% \emph{mean average precision}, 31.07\% \emph{mean precision}, 76.70\% \emph{recall} and 86.15\% \emph{recommendation accuracy}. These results are found promising according to the existing relevant studies \cite{surfclipse, gbing, gyahoo}. We also compare with four popular existing approaches-- \citet{heuristics}, \citet{strathcona}, \citet{selene} and \citet{sourcerer}, for the same dataset, 
and find that our approach outperforms them in all corresponding performance metrics. 
Thus we make the following technical contributions in this paper.
\begin{itemize}
\item We propose a graph-based approach in order to estimate the \emph{structural relevance} between two code segments.
\item In the ranking of exception handling code examples, we not only combine \emph{structural relevance} and \emph{lexical relevance} but also consider the \emph{quality of the exception handlers} in the examples.
\item We package our solution into a tool, \emph{SurfExample} \cite{sep}, that captures the context code in the IDE, and recommends relevant exception handling code examples collecting from a remote web service \cite{sep}, and the service can be leveraged by any IDE.
\end{itemize} 


\section{Motivating Example}\label{sec:motivation}
Let us consider a problem solving scenario, where a developer implements a client module of an Eclipse plugin that accesses a remote web service. Like many other developers, she is concerned about the functional requirements, and uses only a \emph{generic handler} for exception handling (\eg\ highlighted in Listing \ref{lst:socket}). The implementation serves the primary purpose (\eg\ accessing information) of the client module; however, it also poses several threats to future maintenance and evolution of the plugin. First, the generic handler catches all exceptions 
\begin{lstlisting}[label=lst:socket, language=java, escapechar=@, aboveskip=5pt, belowskip=-2em, float=t, frame=bt, caption={Code under Development}]
//more code ...
try {
       URL url=new URL(WEB_SERVICE_URL_WITH_PARAMS);
       HttpURLConnection conn=(HttpURLConnection)url.openConnection();
       //more code goes here ...
    }@\hl{ catch (Exception e) }@{ }// generic exception handler 
\end{lstlisting}
\begin{lstlisting}[label=lst:recommended, language=java,  escapechar=@, aboveskip=5pt, float=t, belowskip=-2em, frame=bt, caption={Recommended Code Example}]
BufferedReader breader=null;	
try {
       URL url = new URL(this.web_service_url);
       HttpURLConnection httpconn = (HttpURLConnection) url.openConnection();
       httpconn.setRequestMethod("GET");
       if (httpconn.getResponseCode() == HttpURLConnection.HTTP_OK) {
               breader = new BufferedReader(new InputStreamReader(
                    httpconn.getInputStream()));
                String line = null;
                while ((line = breader.readLine()) != null) {
		 	 //more code goes here ...
            }
        }
    } catch (MalformedURLException mue) {
      @\hl{Log.warn("Invalid URL " + this.web_service_url, mue);}@
      @\hl{MessageDialog.openError(Display.getDefault()}@.getShells()[0],
                "Invalid URL " + this.web_service_url, mue.getMessage());
    } catch (ProtocolException pe) {
      @\hl{Log.warn("Protocol Exception " + this.web_service_url, pe);}@
      @\hl{MessageDialog.openError(Display.getDefault()}@.getShells()[0],
                "Invalid Protocol " + this.web_service_url, pe.getMessage());
    } catch (IOException ioe) {
      @\hl{Log.warn("Failed to access the data " + this.web_service_url, ioe);}@
    } finally {
      @\hl{breader.close();}@
   }
}
\end{lstlisting}
triggered from within the \emph{try} block and \emph{suppresses} each of them, which clearly violates the second accepted principle \cite{hp2} of exception handling--\emph{if you catch an exception, do not swallow it}. The suppression conceals important information of the occurred exceptions, and identification or fixation of a bug in a multilayer application with such poorly designed handlers is highly error-prone and time-consuming.
Second, exceptions are generally associated with different API methods (according to API design specifications), and several exceptions can occur from a programming context.
Each of those exceptions (especially checked exceptions) deserves a specific treatment (\eg\ handle or rethrow) based on the application context or abstraction.
The generic handler without any cleanup operations fails to meet the individual exception-specific requirements \cite{chang}, and thus leads to different hidden bugs and resource or security issues.

Now let us consider the code example in Listing \ref{lst:recommended} recommended by the proposed approach for the current programming context (\ie\ code under development) in Listing \ref{lst:socket}. The example performs a similar type of programming task using a similar set of API objects, and thus is completely relevant to the current context. The code example treats each of the exceptions that can trigger from the code, and it also provides valuable information for exception handling. 
First, the developer is often not aware of the exceptions which might occur from the current programming context. She also might not be sure of which of the exceptions are to be caught and handled if the IDE suggests them based on API specifications. The recommended relevant example provides such information, and she can easily apply that in the current context. Second, she might also lack necessary skills required to handle the exceptions, and the example demonstrates how certain exceptions should be caught and handled (\eg\ highlighted lines in Listing \ref{lst:recommended}). For example, the technical details of a \emph{MalformedURLException} can be used to warn a user about the URL, and thus it is a candidate exception for handling according to the first principle \cite{hp,hp2} of exception handling--\emph{Always catch only those exceptions that you can actually handle}.
The example handles it through reporting to the user using a dialog box (\ie\ instant notification) and logging the details for future maintenance. In practice, effective handling of exceptions is a frequently misunderstood aspect of programming especially with applications of multilayer abstraction, and such exception handling code examples can act as a helpful learning tool for the developer. The examples are not necessarily meant for reuse; however, they can guide her towards effective handling in her application context through exemplary implementation.
\vspace{-.1cm}
\section{Background}
\label{sec:bg}
\subsection{Static Relationship \& Data Dependency }\label{sec:dependency}
\begin{figure}[!t]
\centering
\noindent\cfbox{gray}{\includegraphics[width=3in]{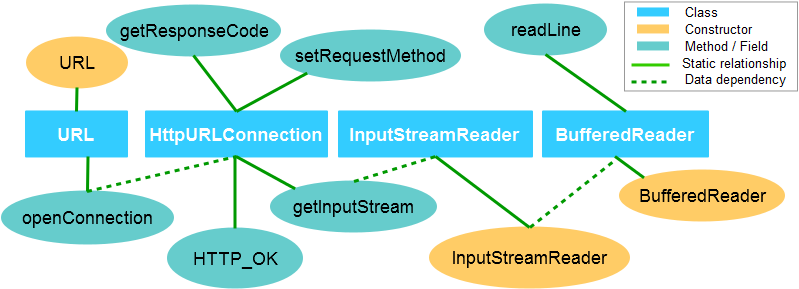}}
\vspace{-.2cm}
\caption{Static Relationship and Data Dependency in Listing \ref{lst:recommended}}
\label{fig:dependency}
\vspace{-.3cm}
\end{figure}

\citet{groum} propose a graph-based approach for \emph{API usage pattern} extraction, where they represent the usage of different API objects in the code using graphs. In the graph, each API object and its properties such as \emph{fields}, \emph{constructors} and \emph{methods} are represented as nodes, and static relationships between the object and its properties or its dependencies on other objects are represented as connecting edges. They classify the dependencies into two-- \emph{data dependency} and \emph{temporal usage order}.
If an API object accepts an instance or an attribute of another object as the parameter either in the constructor or in the method, the first object is said to be dependent on the second object by data. On the other hand, certain methods can be  invoked only after the invocation of another method from the same object. For example, all method invocations of an object are followed by the initialization (\ie\ \emph{<init>} method) of the object, and this type of dependency of sequence is termed as the temporal usage order. 
In this research, we leverage the \emph{static relationships} between API objects and their properties as well as the \emph{data dependencies} among different objects in the code in order to determine the structural relevance between two code segments. Fig. \ref{fig:dependency} shows the static relationships (\ie\ solid edges) and the data dependencies (\ie\ dashed edges) in the recommended code example in Listing \ref{lst:recommended}, which uses four API classes-- \emph{URL, HttpURLConnection, InputStreamReader} and \emph{BufferedReader}. We note that \emph{InputStreamReader} object accesses \emph{getInputStream} method of \emph{HttpURLConnection} object, and \emph{BufferedReader} object accepts an instance of \emph{InputStreamReader} class in their constructors respectively, and these data dependencies are shown using dashed edges. On the other hand, object to property (\eg\ method, constructor) static relationships are represented as green coloured solid edges.

\subsection{Graph Matching}\label{sec:gmatch}
In graph theory, \emph{matching graphs} involves matching a set of independent edges along with their vertices \cite{gmatch}. Fig. \ref{fig:dependency}, the adapted API usage graph of our toy example (Listing \ref{lst:recommended}), shows the static relationships and the data dependencies in the code in terms of vertices and edges. 
In our research, we estimate the \emph{structural relevance} between a candidate code example and the context code, where
we determine matching between two such corresponding usage graphs.
We consider \emph{maximum matching} in the graphs, and 
also estimate different \emph{heuristic weights} for different types of matching (\eg\ dependency, static relations) using a machine-learning based approach (Section \ref{sec:ml}).
For example, a data dependency matching is considered more important (\ie\ contains greater weight) than a static relationship matching for relevance estimation. The static relationship matching between two graphs explains that two graphs merely contain similar set of API objects with their properties (\eg\ method or field). On the other hand, the data dependency matching explains that those API objects also interact with each other in a similar fashion in both graphs, which adds more value in relevance estimation.

\begin{figure}[!t]
\centering
\includegraphics[width=3.5in]{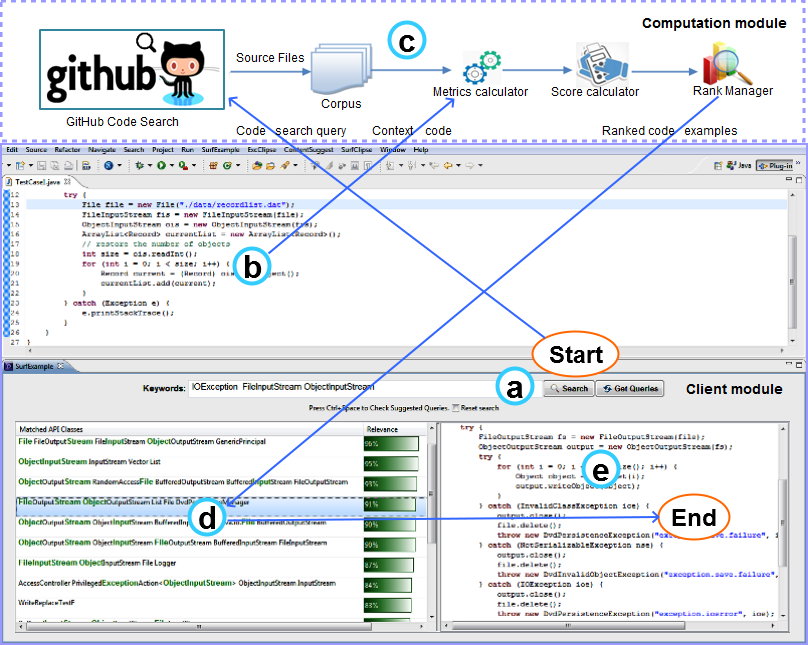}
\vspace{-.6cm}
\caption{Schematic Diagram of Proposed Approach}
\label{fig:sysdiag}
\vspace{-.4cm}
\end{figure}

\section{Proposed Approach}
\label{sec:approach}
Fig. \ref{fig:sysdiag} shows the schematic diagram of our proposed approach for exception handling code example recommendation. We package our solution into an easily accessible web service \cite{sep} and an Eclipse plugin \cite{sep}. In this section, we discuss different working modules of the solution, proposed metrics for relevance estimation between code segments as well as the quality estimation of the exception handlers,
metric weight estimation techniques and ranking algorithms.

\subsection{Working Modules}\label{sec:modules}
The proposed solution adopts a client-server architecture, and it has two working modules-- \emph{client module} (Fig. \ref{fig:sysdiag}-(a, b, d, e)) and \emph{computation module} (Fig. \ref{fig:sysdiag}-(c)). The client module, an Eclipse plugin prototype \cite{sep}, collects the code under development (hereby we call it \emph{context code}) containing generic or poorly designed exception handlers from the IDE, and prepares a search query by extracting suitable keywords from the code (Section \ref{sec:queryform}). It then sends the search query as well as the context code to the computation module.
The computation module, hosted as a web service \cite{sep}, collects code examples from \emph{GitHub code repositories} using that search query and \emph{GitHub code search API}, and develops a dynamic corpus (Fig. \ref{fig:sysdiag}-(c)). The corpus generally contains about 60-70 code examples from hundreds of repositories, which are analyzed, filtered, and then ranked against the context code using the proposed metrics (Section \ref{sec:metrics}) and ranking algorithms (Section \ref{sec:ranking}). Once the ranked examples are returned from the computation module, the client module recommends the top 15 of them in the IDE (Fig. \ref{fig:sysdiag}-(d, e)).
The developer then can check the code examples and leverage for exception handling in her own programming context.

\subsection{Proposed Metrics}\label{sec:metrics}\
This section discusses our proposed metrics which are used to determine the \emph{structural} and \emph{lexical relevance} of a candidate code example in the corpus with the \emph{context code} in the IDE. It also discusses our proposed metrics that estimate the \emph{quality of the exception handlers} in the code example.

\subsubsection{Structural Relevance ($R_{str}$)}
\citet{heuristics} apply heuristic strategies on three structural facts--(1) the hierarchy level of the handled exception, (2) list of methods called, and (3) types of the variables used, of an exception handling code example for structural relevance estimation. \citet{strathcona} also adopt a similar approach to capture the structural information from the code. They develop six heuristic strategies associated with class inheritance, method call and variable usage. Thus, existing two studies \cite{heuristics, strathcona} basically consider number of matched method calls and number of matched variables as the core components of \emph{structural relevance} between two code segments.

In our research, we propose a \emph{graph-based approach} (adapted from the approach of \citet{groum} for \emph{API usage pattern extraction}) to capture the structural features from the code. We consider a code example as a network or graph of API objects interacting with each other through method or constructor invocations and field accesses in order to solve a programming problem. We consider each of the API objects, the static relationships between an API object and its fields or methods, and the data dependencies of the object upon other API objects in the code as the structural features, and we exploit them to estimate the structural relevance (\ie\ structural similarity) between two code segments. Thus the structural relevance is based on four structural aspects-- matched API objects, matched field accesses, matched method calls, and matched data dependencies.

\textbf{API Object Match ($AOM$)}: In the code, different API objects are initialized, and their fields and methods are accessed in order to accomplish a programming task. We use \emph{JavaParser} \cite{javaparser}, an Eclipse AST-based parser, to extract the API objects from the context code and the candidate code examples collected from GitHub. \emph{API Object Match} metric determines the number of matched API objects between the context code and a candidate example. Given that each API object has a predefined set of fields and methods, the metric can be considered as a rough estimate of the functional similarity between the two code fragments.

\textbf{Field Access Match ($FAM$)}: The metric determines the matching between field accesses of an API object in the context code and that of the target object in the candidate code. While existing approaches \cite{heuristics, strathcona} ignore this feature, we use it as a structural component of the code. In practice, the metric accumulates field matching in the candidate code for all API objects in the context code, and indicates the extent to which  both code fragments access the common attributes. 

\textbf{Method Invocation Match ($MIM$)}: We consider method invocations as an important structural component of the code as the API objects generally interact with each other through them. Existing approaches \cite{heuristics, strathcona} do not consider the scope (\ie\ API class instances) of the invoked methods during matching, and thus their method invocation matching might be erroneous (\ie\ same method names are available in different API objects). In our research, we treat each API object as a working unit. We consider the invoked methods from an API object in the context code, and then determine the method invocation match by comparing with the invocation list from the same object in the candidate code example. In practice, the metric considers each API object in the context code and accumulates the invocation match measures.

\textbf{Data Dependency Match ($DDM$)}: The API objects in the code depend on each other for object initialization, method parameters and so on, and we call it \emph{data dependency} among the objects \cite{groum}. We consider the data dependency as a structural component of the code, and we use API usage graph in order to determine the dependency matching. For example, in Fig. \ref{fig:dependency}, the dependencies among the API objects are represented as dashed edges among the vertices.
Given that API libraries are designed with certain dependencies among different API classes, we capture and exploit such dependencies in order to determine the structural relevance between the context code in the IDE and a candidate code example.

We sum up the above four structural components in order to determine the structural relevance score ($R_{str}$) as follows:
\begin{equation}\label{eq:structure}
\setlength\abovedisplayskip{0pt}
\setlength\belowdisplayskip{0pt}
\begin{split}
R_{str}=\alpha \times N+\beta \times \sum_{i=1}^{N}\frac{FAM_{i}}{FAQ_{i}}+\gamma\times \sum_{i=1}^{N}\frac{MIM_{i}}{MIQ_{i}}+\\
\delta \times \sum_{i=1}^{M}DMT_{i} 
\end{split}
\end{equation}
Here, N and M refer to number of matched API objects and number of matched dependencies respectively. $\alpha, \beta, \gamma$ and $\delta$ are the weights of API Object Match (AOM), Field Access Match (FAM), Method Invocation Match (MIM), and Data Dependency Match (DDM) metrics respectively. The weight estimation technique is discussed in Section \ref{sec:ml}. $FAQ_{i}$ and $MIQ_{i}$ are number of field accesses and number of method invocations of an API object from the context code, and $DMT$ refers to the matching weight of each data dependency. For example, if an API object in the candidate code depends on another object through a different access point (\eg\ method, constructor) than that in the \emph{context code} (\ie\ code under development), we call it partial matching (\ie\ weight 0.5). On the other hand, a complete matching (\ie\ weight 1.0) should match both the access points and the target end objects.
\subsubsection{Lexical Relevance ($R_{lex}$)}
While \emph{structural relevance} exploits certain API object-based structural features in the code, \emph{lexical relevance} captures even finer level granularity--\emph{token}. In order to capture token-level relevance between two code fragments and to add more value to relevance estimation, we use two lexical similarity measures-- \emph{cosine similarity} \cite{cos} and \emph{code clone measure}. They also help to overcome the limitations with non-compilable code (\ie\ structural relevance estimation requires the code to be compilable). Cosine similarity focuses on occurrence and frequency of a particular token in the code irrespective of its order, and thus determines the content similarity between two code segments. On the other hand, the code clone measure depends on the clone detection algorithms. 
In the case of cosine similarity calculation, we consider a code fragment as a vector of tokens, and discard insignificant tokens (\eg\ punctuations). We then determine the cosine of the angular distance (\ie\ cosine similarity) between the two such vectors corresponding to the context code and a candidate code example. In case of code clone measure ($S_{ccm}$), we use a state-of-the-art \emph{code clone detection} technique, NiCAD \cite{croy}, where we determine the \emph{longest common subsequence} of tokens between the context code and the candidate code, and then normalize it as follows:
\begin{equation}\label{eq:croy}
\setlength\abovedisplayskip{0pt}
S_{ccm}=\frac{\left | S_{lcs} \right |}{\left |S_{total} \right  |}
\vspace{-.2cm}
\end{equation}
Here, $S_{lcs}$ denotes the longest common subsequence of tokens, and $S_{total}$ denotes the set of tokens extracted from the context code. The measure values between zero to one, and it provides an estimate of the extent to which the candidate code matches with the context code lexically. Thus, the two measures compute the lexical similarity between code from two different viewpoints, and we use them in order to determine the lexical relevance score ($R_{lex}$) as follows:
\begin{equation}
\setlength\abovedisplayskip{1pt}
\setlength\belowdisplayskip{1pt}
R_{lex}=\lambda \times S_{cos}+\sigma \times S_{ccm}
\end{equation}
Here $\lambda$ and $\sigma$ are the weights of the corresponding measures, and they are calculated using a machine learning technique involving logistic regression (Section \ref{sec:ml}).

\subsubsection{Quality of Exception Handler ($Q_{ehc}$)}
The metrics discussed earlier focus on the relevance of an exception handling code example for recommendation; however, \emph{relevance} alone is not sufficient enough for effective recommendation (\ie\ a limitation of existing studies).
The quality of the exception handlers in the code is also an important concern. In addition to the above metrics and measures, we thus also consider the quality of the exception handlers in the code as follows:  

\textbf{Readability ($RA$)}:
Readability of software code refers to a human judgement of how easy the code is to understand \cite{readability}. 
In our research, we consider \emph{readability} as one of most important quality metrics for an exception handler in the code example. The baseline idea is-- \emph{the more readable and understandable the handler code is, the easier it is to leverage in exception handling}. \citet{readability} propose a code readability model trained on human perception of readability and understandability. The model uses different textual features (\eg\ length of identifiers, number of comments, line length) of the code that are likely to affect the human perception of readability. It then predicts a readability score on the scale from zero to one, inclusive, with one describing that the code is highly readable. We use the readily available library \cite{weimertool} by \citeauthor{readability} to calculate the readability metric of the exception handling code examples.

\textbf{Average Handler Actions ($AHA$)}:
The metric calculates the average number of statements (\ie\ actions) in each of the catch clauses in the code example. During calculation, we discard the insignificant statements such as the statements printing stack traces or error messages. We consider the measure as an important indicator of \emph{how extensively (\ie\ meaningfully) data from the caught exceptions are used for handling}. The lower the measure, the poorer the design of the exception handlers. 

\textbf{Handler to Code Ratio ($HCR$)}:
The metric refers to the fraction of the code in the example that is intended for exception handling, and we use \emph{SLOC (Source Lines of Code)} for the calculation.
While the metric indicates the \emph{richness of the code example in handling exceptions}, it also helps to filter out the examples with poorly designed exception handlers (\eg\ generic handler with empty catch block) or long methods. These examples would necessarily contain a large number of program statements compared to the handler statements in the catch clauses, and we use \emph{Handler to Code Ratio} metric to penalize such code examples.

We use the above three quality estimates focusing on distinct aspects, and determine an overall quality estimate for the exception handlers in the code example as follows:
\begin{equation}
\setlength\abovedisplayskip{1pt}
\setlength\belowdisplayskip{1pt}
Q_{ehc}=\mu \times R+\epsilon \times AHA+\kappa \times HCR
\end{equation}
Here, $\mu, \epsilon$ and $\kappa$ are the weights of the corresponding quality metrics, which are calculated using a machine learning technique involving logistic regression (Section \ref{sec:ml}).
While $HCR$ metric is likely to encourage examples with excessive handling code, $AHA$ metric ensures that the handlers contain meaningful statements, and \emph{$RA$} metric penalizes code with too many parentheses \cite{readability} (\ie\ code with too many handlers).

\subsection{Result Scores and Ranking}\label{sec:ranking}
In our research, we consider three important aspects-- \emph{structural relevance, lexical relevance} and \emph{quality of exception handler} for ranking and recommendation of code examples. The \emph{structural relevance} helps to recommend a code example that uses a set of API objects similar to that of the context code in the IDE. Moreover, it ensures that each API object in that set matches with that in the context code in terms of field access, method invocation and data dependency upon other objects. The \emph{lexical relevance} refers to the lexical similarity of a code example against the context code, and it helps to recommend similar type of code examples for possible reuse. The last aspect focuses on the overall quality of the handlers in the code example. It helps to recommend code examples that are highly understandable, and contain good quality handlers for the exceptions of interest. Thus, the \emph{total relevance} ($R_{total}$), for each candidate code example is calculated using the component scores associated with those three aspects in Equation \eqref{eq:total}. The component scores belong to different ranges due to heterogeneous feature values, and each score is \emph{normalized} between zero to one.
\begin{equation}\label{eq:total}
\setlength\abovedisplayskip{1pt}
\setlength\belowdisplayskip{1pt}
R_{total}=w_{str}\times R_{str}+w_{lex}\times R_{lex}+w_{ehc}\times Q_{ehc}
\end{equation}
Here, $R_{str}$, $R_{lex}$ and $Q_{ehc}$ are \emph{structural relevance}, \emph{lexical relevance} and \emph{quality of exception handler} estimates respectively of a candidate code example. $w_{str}, w_{lex}$ and $w_{ehc}$ are the heuristic weights (\ie\ relative importance) of the corresponding metrics, which are calculated using the machine learning approach discussed in Section \ref{sec:ml}. Once we calculate the total scores, we sort the code examples based on their scores, and recommend the top 15 examples to the developer.
For instance, the code example in Listing \ref{lst:recommended} shows these values for the proposed nine individual metrics-- \emph{AOM}=2, \emph{FAM}=0, \emph{MIM}=1, \emph{DDM}=0, \emph{$S_{cos}$}=0.67, \emph{$S_{ccm}$}=0.58, \emph{RA}=0.09, \emph{AHA}=1.67, \emph{HCR}=0.52 during ranking. 
Given the limited (\ie\ a few statements) context code (\ie\ code under development) in Listing \ref{lst:socket}, the example in Listing \ref{lst:recommended} matches with it
the best both structurally and lexically among all other examples. More importantly, exceptions in the example are handled carefully with at least two actions (\ie\ statements) in each handler and the code is moderately readable, and thus
the example gets a normalized handler quality score of  0.68. While other approaches either \emph{analyzes manually} or \emph{depends on the reputation} of the code repository for good quality handlers of exceptions, we not only choose reputed repositories and but also propose and use several metrics to ensure quality of the exception handlers (\ie\ effectiveness shown in Fig. \ref{fig:venn}).
Based on the three aspects (\emph{structual, lexical} and \emph{handler quality}) considered, the example scores the highest, and ranks the top in the recommended example list.

\subsection{Metric Weight Estimation}\label{sec:ml}
In order to determine the weight of \emph{nine} of the individual metrics associated with structural relevance, lexical relevance and handler quality of a code example, we choose 650 code examples handling 65 exceptions from experiment dataset. For each exception, we collect ten random candidate examples from the corpus, analyze their content, and manually tag them either as \emph{relevant} or \emph{irrelevant} for recommendation. We also collect the values of all nine proposed metrics for each tagged code example.
We then feed the feature (\ie\ metric) values and class labels (\ie\ tag of example) to \emph{Weka} tool \cite{weka} that returns a logistic regression based classifier model \cite{logistic}. In the classifier model,  each of the features is associated with certain coefficients, which the tool tunes in order to classify a sample (\ie\ code example) with maximum accuracy.
We believe that these coefficients are an estimate of the importance of the features used in the classification, and we consider them as the weights of the corresponding nine relevance and quality metrics \cite{specmining}. However, the coefficients are either positive (\ie\ supporting for a particular class) or negative (\ie\ discouraging for a particular class), and one may find them counter-intuitive for weight estimates. Therefore, we use \emph{Odd Ratio} of each feature, a logarithmic transformation of the coefficient, as the weight estimate for the corresponding relevance and quality metrics. Among the nine weight estimates, weights of lexical measures dominate others; that means lexical metrics play a decisive role in the classification of the code examples. Weight estimates, and associated data can be found online \cite{sep}.

Once we calculate the subtotal scores using the individual metrics and their corresponding weights, they represent certain aspects such as \emph{structural relevance, lexical relevance} and \emph{exception handler quality} of a code example. We then adopt the same machine-learning technique (as in case of individual metrics above) in order to estimate the relative weights (\ie\ importance) of those three aspects. 
We consider a heuristic relative weight of 1.0152 for \emph{lexical relevance}, 1.2787 for \emph{structural relevance}, and 1.1588 for \emph{exception handler quality} estimate based on the \emph{Odd Ratios} of the corresponding metrics in classifier model.

\section{Experiment}\label{sec:experiment}

\subsection{Dataset Preparation}\label{sec:dataset}
We collect 65 exception handling cases (\ie\ scenarios) for the experiments, where each case comprises of a \emph{context code segment} and an exception to be handled.
Most of the cases are collected from different online sources such as Pastebin \cite{pb} and StackOverflow Q \& A site, and a few of them are developed by us.
For each of the cases, the context code is analyzed to prepare a suitable search query (Section \ref{sec:queryform}), which is then used to develop a corpus of
candidate code examples containing handlers of the corresponding exception.
In order to collect examples, we choose four popular software organizations--\emph{Apache, Eclipse, Facebook} and \emph{Twitter}, and they host about 738 open source Java projects (visited on January, 2014) at GitHub. The code bases of the target organizations are considerably rich and matured, and some of the organizations even developed exception handling frameworks (\eg\ ExceptionUtils and Camel by Apache). Thus we believe that their code bases are more likely to contain code examples with efficient handlers for exceptions. 
We use \emph{GitHub Code Search} and the prepared search queries to collect the code examples. For each of the cases, we collect 60-70 candidate code examples containing exception handlers, and the whole corpus contains about 4,400 examples in total.


\subsection{Search Query Formulation}\label{sec:queryform}
During corpus development, we prepare a search query for each of the exception handling cases, and collect the candidate code examples from GitHub code search using that query.
Each of those queries generally contains two types of information--\emph{exception name} and \emph{dominant API class name}. We analyze the context code to extract such information, where we experience two exception handling scenarios. In the first scenario, the context code specifies which exception to be handled, and we use that exception name in the search query. In the second scenario, the context code either does not specify the exception or contains a generic exception handler (\eg\ Listing \ref{lst:socket}), and we adopt a careful approach to choose an exception (to be handled) for this scenario. Given that exceptions are associated with different API methods (according to API design specifications), we consider all the checked exceptions those might be thrown from within the context code, and choose the one that is the most frequent with the API methods in the code.
In case of dominant API class name token in the search query, we analyze the API objects used in the context code. 
The idea is to identify the most active API objects in the code, and we consider an object with the most frequent method invocation and field access as the most active API object.
Thus the search query for the context code in Listing \ref{lst:socket} is-- \emph{IOException URL}.

\subsection{Exception Oracle Development}\label{sec:oracle}
We develop an \emph{oracle} that returns a list of the most relevant code examples for each of the exception handling cases. For oracle development, we analyze code examples in the corpus collected for each case, and 
check for their relevance against the corresponding context code and the exception of interest.
Given that checking the relevance of a code example against an exception and its context code is a subjective approach, 
and a number of examples are associated with each case, we use tool support in our analysis. First, we rank the examples based on their lexical similarity against the context code, and then manually check them from the top for relevance.
We consult the best accepted practices \cite{hp} for exception handling, look for meaningful actions (\eg\ cleanup, rethrow, status notification) other than logging in the exception handlers of a code example, and use our best judgement to choose the relevant examples. Once the examples are chosen for the oracle, they are cross-validated by the peers (\eg\ two graduate research students with at least five years of Java programming experience), and we finalize the example list through discussion.
We choose 176 code examples as the most relevant ones for 65 exception handling cases. It took about 50-60 working hours. The code examples are hosted online \cite{sep}, and we use them as the benchmark examples to determine the performance of the proposed and existing approaches.

\subsection{Performance Metrics}\label{sec:permet}
Our proposed approach profoundly aligns with the research areas of information retrieval and recommendation systems. We thus use a list of performance metrics for evaluation from those areas as follows:

\textbf{Mean Precision (MP)}: \emph{Precision} determines the percentage of the results (\ie\ code examples) returned by a query (\ie\ exception handling case) that is relevant.
\emph{Mean Precision} averages that percentage for all queries in the dataset.

\textbf{Mean Average Precision at K (MAPK)}: \emph{Precision at K} calculates \emph{precision} at the occurrence of every relevant result (\ie\ relevant code example) in the ranked list. \emph{Average Precision at K (APK)} averages the \emph{precision at K} for all relevant results in the list for a query (\ie\ exception handling case). \emph{Mean Average Precision} is the mean of \emph{average precision at K} for all queries in the dataset.



\textbf{Recall (R)}: \emph{Recall} denotes the fraction of all the relevant results (\ie\ benchmark examples) that are retrieved. 
\subsection{Experimental Results}
We conduct experiments with 65 exceptions (related to standard Java development) along with their context code segments,
and collect the top 15 recommended code examples for each of the exceptions for evaluation. We analyze the results and determine the performance using necessary metrics (Section \ref{sec:permet}).
This section discusses the experimental results and the recommendation performance of our approach.

\begin{table}[!t]
\caption{ Experimental  Results}
\label{table:proposed}
\centering
\resizebox{2in}{!}{%
\begin{threeparttable}[b]
\begin{tabular}{l|c|c|c}
\hline
 \textbf{Metric} & \textbf{Top 5} & \textbf{Top 10} & \textbf{Top 15}\\
\hline
MP & \textbf{31.07\%} & 18.62\% & 13.85\% \\
\hline
MAPK& \textbf{41.92\%} & 39.92\% & 38.64\%\\
\hline
TEH\tnote{1} (65) & 48(101) & 53(121) &\textbf{ 56}(135) \\
\hline
PEH\tnote{2}&73.85\% & \textbf{81.54\%} & \textbf{86.15\%}\\
\hline
R\tnote{3} & 57.39\% & 68.75\% & \textbf{76.70\%}\\
\hline
\end{tabular}
$^1$No. of exceptions handled, $^2$\% of all exceptions handled, $^3$\% of relevant examples recommended
\end{threeparttable}
}
\vspace{-.2cm}
\end{table}
\begin{figure}[!t]
\centering
\includegraphics[scale=.35]{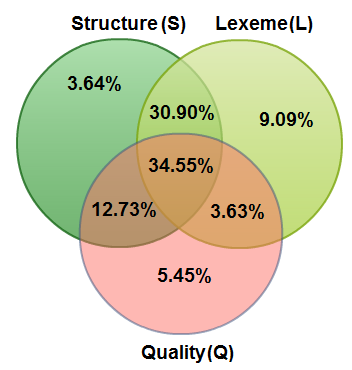}
\vspace{-.3cm}
\caption{Result Distribution over Metrics}
\label{fig:venn}
\vspace{-.5cm}
\end{figure}
\begin{table*}[!t]
\caption{ Experimental Results on Different Score Components}

\label{table:component}
\centering
\resizebox{6in}{!}{%
\begin{threeparttable}[b]
\begin{tabular}{l|l|c|c|c||l|l|c|c|c}
\hline
\textbf{Single aspect}& \textbf{Metric} & \textbf{Top 5} & \textbf{Top 10} & \textbf{Top 15}&\textbf{Combined aspects}& \textbf{Metric} & \textbf{Top 5} & \textbf{Top 10} & \textbf{Top 15}\\
\hline
\multirow{5}{*}{Structure ($R_{str}$)}&MP & 27.07\% &16.76\% & 12.51\%& \multirow{3}{*}{Structure ($R_{str}$),} & MP & 27.99\% & 17.99\% & 13.44\%\\
& MAPK & 38.07\% & 33.84\% & 32.64\%& &MAPK & 43.08\% & 38.69\% & 37.33\%\\
& TEH(65)  & 45(88) & 49(109) & 53(122) &\multirow{2}{*}{and Content ($R_{lex}$)} &TEH(65)& 45(91) & 49(117) & 53(131) \\
&PEH & 69.23\% & 75.38\% & 81.54\% & &PEH & 69.23\% & 75.38\% & 81.54\%\\
& R& 50.00\% & 61.93\% & 69.32\% & &R & 51.70\% & 66.48\% & \textbf{74.43\%} \\
\hline
\multirow{5}{*}{Content ($R_{lex}$)} & MP & 24.62\% & 17.23\% & 12.72\%&\multirow{3}{*}{Structure ($R_{str}$), } & MP & \textbf{31.07\%} & 18.62\% & 13.85\%\\
&MAPK & 35.00\% & 33.85\% & 33.08\%& &MAPK& \textbf{41.92\%} & 39.92\% & 38.64\%\\
&TEH(65) & 43(80) & 49(112) & 53(124) &Content ($R_{lex}$), & TEH(65) & \textbf{48}(101) & \textbf{53}(121) &\textbf{ 56}(\textbf{135})\\
&PEH& 66.15\% &75.38\% & 81.54\% &and Quality ($Q_{ehc}$ ) &PEH & 73.85\% & 81.54\% & \textbf{86.15\%}\\
&R & 45.45\% & 63.63\% & 70.45\% & & R & 57.39\% & 68.75\% & \textbf{76.70\%}\\
\hline
\end{tabular}
\textbf{TEH}=Total exceptions handled, \textbf{PEH}=Percentage of all exceptions handled
\end{threeparttable}
}
\vspace{-.5cm}
\end{table*}
Table \ref{table:proposed} shows the results of the experiments conducted on the proposed approach, where we apply different performance metrics such as \emph{Mean Precision (MP), Mean Average Precision at K (MAPK), Total Exceptions Handled (TEH), Percentage of all Exceptions Handled  (PEH)} and overall \emph{Recall (R)}. We collect the top 5, top 10 and top 15 code examples from the recommendation list for evaluation. From Table \ref{table:proposed}, we note that the approach provides results with 31.07\% \emph{mean precision}. That means, on average the technique recommends  31.07\% relevant code examples for each of the exception handling cases, and it recommends correctly for 86.15\% of the exceptions. It also successfully recommends 135 out of 176 benchmark relevant examples, which gives an over all \emph{recall} of 76.70\%. More interestingly, our approach recommends relevant code examples for 48 (73.85\%) out of 65 exceptions with 41.92\% \emph{mean average precision} even when only top 5 results are considered. These results are also found promising according to relevant existing studies \cite{surfclipse, gbing, gyahoo}.

Fig. \ref{fig:venn} shows the distribution of the \emph{handled} (\ie\ code examples correctly recommended) exceptions over different metrics--\emph{structural relevance (S)}, \emph{lexical relevance (L)} and \emph{exception handler quality (Q)}. 
The distribution over a metric means that a certain fraction of the exceptions are handled (\ie\ relevant code examples recommended) considering that metric in isolation.
We note that the handled exceptions are largely distributed over \emph{structural} and \emph{lexical relevance} metrics compared to \emph{exception handler quality}, and all three metrics share about 34.55\% of the exceptions. More interestingly, we note that about 18\% (from Fig. \ref{fig:venn}, 3.64\% + 9.09\% + 5.45\%) exception handling cases are unique to the three metrics, which indicates that those exceptions cannot
be handled or relevant code examples cannot be retrieved without considering those metrics in combination.

Table \ref{table:component} further motivates the idea of \emph{combined relevance} and \emph{quality measures} with statistical evidences. It shows the results of the experiments, where we contrast among the three aspects of relevance and exception handler quality of the code examples. From Table \ref{table:component}, we note that the different relevance aspects such as \emph{lexical relevance} and \emph{structural relevance} are not satisfactorily effective especially in terms of \emph{mean average precision} and \emph{recall}, when they are considered in isolation. For example, the approach can recommend at most 70.45\% of the relevant code examples with 35.00\% \emph{mean average precision} when we consider only \emph{lexical relevance} for ranking. On the other hand, when we consider both \emph{structural} and \emph{lexical relevance}, the approach can recommend 
with 74.43\% \emph{recall} and 37.33\% \emph{precision}. One can argue that performance improvement is not significant, which actually motivates the inclusion of another dimension in code example ranking. We consider \emph{quality of exception handler} as the third aspect in the relevance ranking of the code examples, and we also find it promising in our experiments. When we add \emph{handler quality} to the rest two aspects of ranking, we get a maximum \emph{recall} of 76.70\% and \emph{mean average precision} of 41.92\% by the proposed approach, and it also handles a maximum of 86.15\% of all the exceptions in the dataset. While the improvement is not still too high, the combination of three aspects interestingly performs the best in terms of all performance metrics, and the results are promising. Similar findings can also be reported from Fig. \ref{fig:venn}.  
\vspace{-.1cm}
\subsection{Comparison with Existing Approaches}\label{sec:existing}
Even though our proposed approach shows promise in the controlled experiments above, we further wanted to see how good the approach is in terms of the literature. Thus, we compare our approach with four well known existing approaches-- \citet{heuristics}, \citet{strathcona}, \citet{selene} and \citet{sourcerer}. We implemented the approaches in our working environment based on the methodologies described in the paper and our prior development experience, tested with our dataset, and analyzed their performance with the same set of metrics. This section discusses the comparative study between our proposed approach and the existing approaches.

\citet{heuristics} developed their corpus by collecting code examples from the repositories hosted at \emph{Eclipse Foundation Open Source Community}. They apply different preprocessing on the examples such as discarding inefficient handlers and long methods and so on, and they then apply three heuristics related to \emph{exception type, method call} and \emph{variable usage} for the relevance ranking. In our implementation of the approach,
although we could not replicate their preprocessing steps properly, we used our example corpus as the dataset, and implemented their heuristics according to the guidelines described in the paper.
We thus basically compare our proposed metrics with their proposed heuristics in terms of different experiments.
Table \ref{table:existing} shows the findings of the comparative study, where we observe that their heuristic-based approach performs relatively poor in recommendation. The approach by \citeauthor{heuristics} recommends relevant code examples at most for 44.62\% of the exceptions with 31.25\% \emph{recall} and 16.15\% \emph{mean average precision}, whereas our approach can recommend for 86.15\% of the exceptions with 76.70\% \emph{recall} and 41.92\% \emph{mean average precision}. This clearly shows that our approach outperforms their approach. One can rationalize the lack of preprocessing for the low performance of their approach, we argue that the same limitation is also acknowledged by \citeauthor{heuristics}, and this actually validates that our proposed metrics are more effective than their heuristics for the recommendation from the same corpus. 

Although the remaining three approaches are not especially designed for recommending exception handling code examples, they are well known code example recommendation techniques and are closely related to our work. They also analyze either structural or lexical features from the code for recommendation, and we compare our approach against them. We implemented the existing approaches with required adjustments for the comparative study as the implementations by the authors are either unavailable or not directly applicable. The approach by \citet{strathcona} uses six heuristics for code recommendation, and we find three of them are relevant for exception handling code recommendation. We thus use the three heuristics dealing with \emph{method calls} and \emph{variable usages} in the code. 
\citet{selene} use \emph{cosine similarity} in order to determine relevance between two code examples. \citet{sourcerer} adopt an information retrieval-based approach for code example recommendation. They extract the tokens containing different structural information from the code, and develop a \emph{lucene index} for all the examples in the corpus. They then use a structured query containing a set of predefined parameters to collect recommendable code examples. In our implementation, we adopt a similar approach in index development involving \emph{lucene indexer}; however, we follow a different approach for query formulation. Their query parameters \cite {sourcerer} are not sufficient enough to request for exception handling code examples, and we use the queries (Section \ref{sec:queryform}) by our proposed approach. However, as the experiment results suggest, none of the three existing approaches perform considerably well in recommending exception handling code examples. From Table \ref{table:existing}, we note that the approach by \citeauthor{selene} \emph{handles} a maximum of 31 (47.69\%) of exceptions and recommends examples with 21.54\% \emph{mean average precision} and 30.68\% \emph{recall}, and others recommend less than 30\% of all the relevant examples (\ie\ \emph{recall}), which are significantly poor compared to our results. One can argue that the comparison might not be fair due to the \emph{handler quality metrics} in our approach. However, as shown in Table \ref{table:component}, our approach also performs significantly better than those approaches without using those metrics.
Thus, we conjecture that those approaches were not actually designed for exceptional handling code recommendation; but to the best of our knowledge they are worthy of comparison as there are no others available. 
\begin{figure}[!t]
\centering
\includegraphics[width=2.5in]{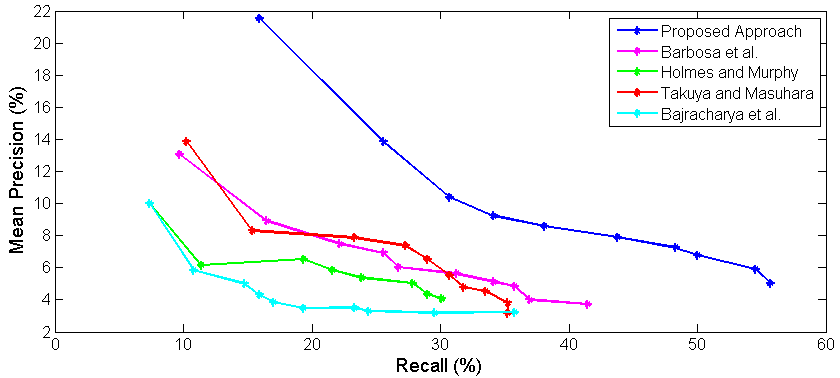}
\vspace{-.4cm}
\caption{Mean Precision vs. Recall Curves}
\label{fig:mpr}
\vspace{-.6cm}
\end{figure}

As shown in the schematic diagram in Fig. \ref{fig:sysdiag}-(c), our approach leverages GitHub code search in dynamic corpus development. The approach thus applies ranking algorithms on a narrowed-down dataset for each exception handling case, whereas other approaches deal with a large local or remote corpus for the same. We also investigate if this additional search (\ie\ GitHub search) is the sole factor behind the promising results of our approach, and conduct experiments with 4,400 code examples as the corpus for each of the exception handling cases. From Table \ref{table:existing}, we note that our approach also performs significantly well compared to the existing approaches in this case. It recommends relevant code examples for a maximum of 40 (61.54\% compared to 47.69\% of existing approaches) out of 65 exception handling cases with 43.75\% \emph{recall}. The \emph{mean precision-recall} curve in Fig. \ref{fig:mpr} also shows that the proposed approach is more promising than the existing approaches in exception handling code recommendation. Given that the area under the curve denotes the performance of a system, our approach outperforms all other approaches in the experiments.
\begin{table*}[!t]
\caption{ Comparison with Existing Approaches}
\label{table:existing}
\centering
\resizebox{6.6in}{!}{%
\begin{threeparttable}[b]
\begin{tabular}{l|l|c|c|c||l|l|c|c|c}
\hline
\textbf{Recommender}& \textbf{Metric} & \textbf{Top 5} & \textbf{Top 10} & \textbf{Top 15} & \textbf{Recommender}& \textbf{Metric} & \textbf{Top 5} & \textbf{Top 10} & \textbf{Top 15}\\
\hline

\multirow{5}{*}{\citet{heuristics}}&MP & 8.92\% &6.92\% & 5.64\% & \multirow{3}{*}{Proposed Approach (without GitHub search)} & MP & \textbf{13.54\%} & 8.77\% & 7.18\%\\
& MAPK & \textbf{16.15\%} & 14.69\% & 13.72\%& &MAPK & \textbf{21.80\%} & 19.87\% & 18.85\%\\
& TEH(65)  & 18(29) & 25(45) & 29(55) &\multirow{2}{*}{Structure($R_{str}$) only} &TEH(65)& \textbf{30}(44) & \textbf{33}(57) & \textbf{37}(70)\\
&PEH &27.69\% & 38.46\% & \textbf{44.62\%} & &PEH & 46.15\% & 50.77\% & 56.92\%\\
& R& 16.47\% & 25.57\% & \textbf{31.25\%}& &R & \textbf{25.00\%} & \textbf{32.38\%} & \textbf{39.77\%} \\
\hline

\multirow{5}{*}{\citet{strathcona}} & MP & 6.15\% & 5.85\% & 5.03\% & \multirow{3}{*}{Proposed Approach (without GitHub search)} & MP & \textbf{13.85\%} & 9.23\% & 7.90\%\\
&MAPK & 4.62\% & 2.31\% & 2.31\%& &MAPK & \textbf{30.64\%} & 27.44\% & 25.90\%\\
&TEH(65) & 16(20) & 25(38) & 31(49) & Structure($R_{str}$), Content($R_{lex}$), &TEH(65)& \textbf{31}(45) & \textbf{34}(60) & \textbf{40}(77)\\
&PEH & 24.62\% &38.46\% & \textbf{47.69\%} & and Quality($Q_{ehc}$) &PEH & \textbf{47.69\%} & 52.31\% & \textbf{61.54\%}\\
&R & 11.36\% & 21.59\% & \textbf{27.84\%} & &R & \textbf{25.56\%} & \textbf{34.09\%} & \textbf{43.75\%}\\
\hline

\multirow{5}{*}{\citet{selene}} & MP & 8.31\% & 7.38\% & 5.54\% & \multirow{3}{*}{Proposed Approach (with GitHub search)} & MP & \textbf{27.99\%} & 17.99\% & 13.44\%\\
&MAPK & \textbf{21.54\%} & 20.51\% & 19.74\% & &MAPK & \textbf{43.07\%} & 38.69\% & 37.33\%\\
&TEH(65) & 22(27) & 31(48) & 31(54) & \multirow{2}{*}{Structure($R_{str}$) and Content($R_{lex}$)} &TEH(65)& 45(91) & 49(117) & \textbf{53}(131)\\
&PEH & 33.85\% & 47.69\% & \textbf{47.69\%} & &PEH&69.23\% & 75.38\% & \textbf{81.54\%}\\
&R & 15.34\% & 27.27\% & \textbf{30.68\%} & &R & \textbf{51.70\%} & \textbf{66.48\%} & \textbf{74.43\%}\\
\hline

\multirow{5}{*}{\citet{sourcerer}} & MP & 5.85\% & 4.31\% & 3.49\% & \multirow{3}{*}{Proposed Approach (with GitHub search)} & MP & \textbf{31.07\%} & 18.62\% & 13.85\%\\
&MAPK & 8.46\% & 7.95\% & 6.41\% & &MAPK & \textbf{41.92\%} & 39.92\% & 38.64\%\\
&TEH(65) & 12(19) & 18(28) & 20(34)&Structure($R_{str}$), Content($R_{lex}$), &TEH(65)& \textbf{48}(101) & \textbf{53}(121) & \textbf{56}(135)\\
&PEH & 18.46\% & 27.69\% & \textbf{30.77\%}&and Quality($Q_{ehc}$) &PEH&73.85\% & 81.54\% & \textbf{86.15\%}\\
&R & 10.80\% & 15.91\% & \textbf{19.32\%}& &R & \textbf{57.39\%} & \textbf{68.75\%} & \textbf{76.70\%}\\
\hline
%
%
\end{tabular}
\textbf{TEH}=Total exceptions handled, \textbf{PEH}=Percentage of all exceptions handled
\end{threeparttable}
}
\vspace{-.5cm}
\end{table*}
\vspace{-.4cm}
\section{Threats to Validity}
\label{sec:threats}
In our proposed approach, we note several issues worthy of discussion. First, one might argue about the reliability of the judges for the oracle, especially because relevance checking of a code example against an exception (and its context code) is a subjective approach. In order to overcome this threat, we carefully chose the examples by consulting the best accepted practices of exception handling as well as based on our best judgment, and both authors have professional development experience (details in Section \ref{sec:oracle}).

Second, we exploit \emph{GitHub code search API} to develop the corpus for our experiments, and our approach is subjected to strengths and weaknesses of the search feature. One might argue about the relatively smaller size of the corpus developed dynamically for each of the exception handling cases.
However, we argue that those examples are actually collected from hundreds of open source repositories (about 750), and then filtered and even ranked before returning. Thus, the developed corpus was not only sufficient for our experiments but also an effective one, which is also shown by the experimental results.

Third, one might argue about the number of exceptions for the experiments. We used 65 exception handling cases for the experiments and this might not be sufficient enough to draw a generalized conclusion. However, collecting suitable cases and developing reliable oracle for them requires significant amount of time and efforts, and we covered most of the well known standard Java exceptions \cite{sep} in different cases. The corpus is also developed using examples from hundreds of code repositories hosted online. Thus we believe that the sample size is sufficient enough for a controlled experiment and to draw such a conclusion.
\vspace{-.1cm}
\section{Related Work}
\label{sec:related}
Exception handling is not a new topic, and there exists a good number of studies \cite{robi, goodenough, chang, fieldstudy, garcia, heuristics}.
\citet{heuristics} propose an approach to recommend exception handling code by exploiting three heuristics about structural facts in the code. 
The approaches by \citet{strathcona, selene} and \citet{sourcerer} are well known as code recommendation techniques although they are not specialized for exception handling code. 
We compared our approach to all four of them and found that ours one performs significantly better than all of them. For a detailed comparison the readers are referred to Section \ref{sec:existing}. 

The other existing studies on exception handling are not directly related to code example recommendation, and thus, they were not applicable for the comparison experiments. \citet{chang} propose a static analysis technique that considers the exceptional control flows, and helps to discard unnecessary \emph{try-catch} and \emph{throw} statements. However, discarding unnecessary elements from the code may not always meet the needs of the developer in exception handling. \citet{robi} propose another static analysis approach that identifies different possible exceptional flows in the application program and helps the developer to understand and improve the exception handling structures of the system. However, it returns thousands of possible control flow paths for an exception, and that information is not easy to use in practical sense \cite{heuristics}. Thus, in general, the static analysis-based techniques provide limited support for instant exception handling from the first place, and they often assume that handling is already done somehow and the handler code is there \cite{heuristics}. \citet{garcia} conduct an empirical study on the exception handling mechanisms available in different object-oriented programming languages, and propose a new exception handling structure that considers 10 important aspects related to handling. 
\citet{shah} propose a visualization approach that visualizes the exception handling structures in the large software systems for better understanding of how the system works.
Thus while other studies provide useful insights into the control flows, handling structures through static analysis, field studies, empirical studies and visualization, our proposed approach provides readily available and relevant working code examples by exploiting context code in the IDE, which can be easily leveraged for exception handling.
\balance
\section{Conclusion \& Future Works}\label{sec:conclusion}
To summarize, we propose a context-aware code recommender that recommends exception handling code examples against the code under development (\ie\ context code) in the IDE.
We consider three aspects--\emph{structure, content} and \emph{handler quality} of the candidate code examples for relevance ranking.
Experiments with 65 exceptions (and their context code) and 4,400 code examples as well as
comparisons with four existing approaches show that our approach is highly promising.  
While our experiments show that the general-purpose code recommendation approaches are not satisfactorily applicable for the recommendation of exception handling code, in this paper, our technical contribution lies in proposing a graph-based approach for structural relevance estimation, introducing handler quality dimension in relevance ranking, and developing an Eclipse plugin.
In future, we plan to conduct a user study with prospective participants.

\bibliographystyle{plainnat}
\scriptsize
\setlength{\bibsep}{0.0pt}
\bibliography{sigproc}

\begin{thebibliography}{28}
\providecommand{\natexlab}[1]{#1}
\providecommand{\url}[1]{\texttt{#1}}
\expandafter\ifx\csname urlstyle\endcsname\relax
  \providecommand{\doi}[1]{doi: #1}\else
  \providecommand{\doi}{doi: \begingroup \urlstyle{rm}\Url}\fi

\bibitem[cos()]{cos}
Cosine {S}imilarity.
\newblock URL \url{http://en.wikipedia.org/wiki/Cosine_similarity}.

\bibitem[git()]{gitapi}
Git{H}ub {C}ode {S}earch.
\newblock URL \url{http://developer.github.com/v3/search/}.

\bibitem[gma()]{gmatch}
{G}raph {M}atching.
\newblock URL \url{http://en.wikipedia.org/wiki/Matching_(graph_theory)}.

\bibitem[hp()]{hp}
Exception {H}andling {P}rinciples.
\newblock URL
  \url{http://howtodoinjava.com/2013/04/04/java-exception-handling-best-practices}.

\bibitem[hp2()]{hp2}
Best {P}ractices for {E}xception {H}andling.
\newblock URL \url{https://www.ibm.com/developerworks/library/j-ejbexcept}.

\bibitem[jav()]{javaparser}
{J}avaparser-{J}ava 1.5 {P}arser and {AST}.
\newblock URL \url{http://code.google.com/p/javaparser}.

\bibitem[log()]{logistic}
Logistic {R}egression.
\newblock URL \url{http://en.wikipedia.org/wiki/Logistic_regression}.

\bibitem[pb()]{pb}
Pastebin.
\newblock URL \url{http://pastebin.com}.

\bibitem[sep()]{sep}
Surf{E}xample {P}ortal.
\newblock URL \url{http://www.usask.ca/~mor543/surfexample}.

\bibitem[wei()]{weimertool}
Readability {L}ibrary.
\newblock URL \url{http://www.arrestedcomputing.com/readability}.

\bibitem[wek()]{weka}
Weka.
\newblock URL \url{http://www.cs.waikato.ac.nz/ml/weka/}.

\bibitem[Bajracharya et~al.(2009)Bajracharya, Ossher, and Lopes]{sourcerer}
S.~Bajracharya, J.~Ossher, and C.~Lopes.
\newblock Sourcerer: {A}n {I}nternet-{S}cale {S}oftware {R}epository.
\newblock In \emph{Proc. SUITE}, pages 1--4, 2009.

\bibitem[Barbosa et~al.(2012)Barbosa, Garcia, and Mezini]{heuristics}
E.~A. Barbosa, A.~Garcia, and M.~Mezini.
\newblock Heuristic {S}trategies for {R}ecommendation of {E}xception {H}andling
  {C}ode.
\newblock In \emph{Proc. SBES}, pages 171--180, 2012.

\bibitem[Buse and Weimer(2010)]{readability}
R.~P.~L. Buse and W.~R. Weimer.
\newblock {L}earning a {M}etric for {C}ode {R}eadability.
\newblock \emph{TSE}, 36\penalty0 (4):\penalty0 546--558, 2010.

\bibitem[Cabral and Marques(2007)]{fieldstudy}
B.~Cabral and P.~Marques.
\newblock {E}xception {H}andling: {A} {F}ield {S}tudy in {J}ava and .{NET}.
\newblock In \emph{Proc. ECOOP}, pages 151--175, 2007.

\bibitem[Chang et~al.(2001)Chang, Jo, Yi, and Choe]{chang}
B.~M. Chang, J.~W. Jo, K.~Yi, and K.~M. Choe.
\newblock Interprocedural {E}xception {A}nalysis for {J}ava.
\newblock In \emph{Proc. SAC}, pages 620--625, 2001.

\bibitem[Garcia et~al.(2001)Garcia, Rubira, Romanovsky, and Xu]{garcia}
A.~F. Garcia, C.~M.~F. Rubira, A.~Romanovsky, and J.~Xu.
\newblock A {C}omparative {S}tudy of {E}xception {H}andling {M}echanisms for
  {B}uilding {D}ependable {O}bject-{O}riented {S}oftware.
\newblock \emph{JSS}, 59\penalty0 (2):\penalty0 197--222, 2001.

\bibitem[Goodenough(1975)]{goodenough}
J.~B. Goodenough.
\newblock {E}xception {H}andling: {I}ssues and a {P}roposed {N}otation.
\newblock \emph{Commun. ACM}, 18\penalty0 (12):\penalty0 683--696, 1975.

\bibitem[Holmes and Murphy(2005)]{strathcona}
R.~Holmes and G.~C. Murphy.
\newblock Using {S}tructural {C}ontext to {R}ecommend {S}ource {C}ode
  {E}xamples.
\newblock In \emph{Proc. ICSE}, pages 117--125, 2005.

\bibitem[Kumar and Prakash(2009)]{gyahoo}
B.~T.~S. Kumar and J.~N. Prakash.
\newblock {P}recision and {R}elative {R}ecall of {S}earch {E}ngines: {A}
  {C}omparative {S}tudy of {G}oogle and {Y}ahoo.
\newblock \emph{J. Lib. and Info. Mgmt.}, 38\penalty0 (1):\penalty0 124--137,
  2009.

\bibitem[Le~Goues and Weimer(2012)]{specmining}
C.~Le~Goues and W.~Weimer.
\newblock {M}easuring {C}ode {Q}uality to {I}mprove {S}pecification {M}ining.
\newblock \emph{TSE}, 38\penalty0 (1):\penalty0 175--190, 2012.

\bibitem[Nguyen et~al.(2009)Nguyen, Nguyen, Pham, Al-Kofahi, and Nguyen]{groum}
T.~T. Nguyen, H.~A. Nguyen, N.~H. Pham, J.~M. Al-Kofahi, and T.~N. Nguyen.
\newblock Graph-based {M}ining of {M}ultiple {O}bject {U}sage {P}atterns.
\newblock In \emph{Proc. ESEC/FSE}, pages 383--392, 2009.

\bibitem[Rahman et~al.(2014)Rahman, Yeasmin, and Roy]{surfclipse}
M.~M. Rahman, S.~Yeasmin, and C.~K. Roy.
\newblock Towards a {C}ontext-{A}ware {IDE}-{B}ased {M}eta {S}earch {E}ngine
  for {R}ecommendation about {P}rogramming {E}rrors and {E}xceptions.
\newblock In \emph{Proc. CSMR-WCRE}, pages 194--203, 2014.

\bibitem[Robillard and Murphy(2003)]{robi}
M.~P. Robillard and G.~C. Murphy.
\newblock {S}tatic {A}nalysis to {S}upport the {E}volution of {E}xception
  {S}tructure in {O}bject-{O}riented {S}ystems.
\newblock \emph{TOSEM}, 12\penalty0 (2):\penalty0 191--221, 2003.

\bibitem[Roy and Cordy(2008)]{croy}
C.~K. Roy and J.~R. Cordy.
\newblock {NICAD}: {A}ccurate {D}etection of {N}ear-{M}iss {I}ntentional
  {C}lones {U}sing {F}lexible {P}retty-{P}rinting and {C}ode {N}ormalization.
\newblock In \emph{Proc. ICPC}, pages 172--181, 2008.

\bibitem[Shah et~al.(2008)Shah, G\"{o}rg, and Harrold]{shah}
H.~Shah, C.~G\"{o}rg, and M.~J. Harrold.
\newblock {V}isualization of {E}xception {H}andling {C}onstructs to {S}upport
  {P}rogram {U}nderstanding.
\newblock In \emph{Proc. SoftVis}, pages 19--28, 2008.

\bibitem[Takuya and Masuhara(2011)]{selene}
W.~Takuya and H.~Masuhara.
\newblock A {S}pontaneous {C}ode {R}ecommendation {T}ool {B}ased on
  {A}ssociative {S}earch.
\newblock In \emph{Proc. SUITE}, pages 17--20, 2011.

\bibitem[Usmani et~al.(2012)Usmani, Pant, and Bhatt]{gbing}
T.~Usmani, D.~Pant, and A.~K. Bhatt.
\newblock A {C}omparative {S}tudy of {G}oogle and {B}ing {S}earch {E}ngines in
  {C}ontext of {P}recision and {R}elative {R}ecall {P}arameter.
\newblock \emph{J. CSE}, 4\penalty0 (1):\penalty0 21--34, 2012.

\end{thebibliography}


\begin{thebibliography}{20}
\providecommand{\natexlab}[1]{#1}
\providecommand{\url}[1]{\texttt{#1}}
\expandafter\ifx\csname urlstyle\endcsname\relax
  \providecommand{\doi}[1]{doi: #1}\else
  \providecommand{\doi}{doi: \begingroup \urlstyle{rm}\Url}\fi

\bibitem[jav()]{javaparser}
Javaparser-java 1.5 parser and ast.
\newblock URL \url{http://code.google.com/p/javaparser}.

\bibitem[se()]{se}
Surfexamples experiment data.
\newblock URL \url{http://www.usask.ca/~mor543/surfexample}.

\bibitem[gma(Visited on January, 2014)]{gmatch}
Graph matching, Visited on January, 2014.
\newblock URL \url{http://en.wikipedia.org/wiki/Matching_(graph_theory)}.

\bibitem[Bajracharya et~al.(2009)Bajracharya, Ossher, and Lopes]{sourcerer}
S.~Bajracharya, J.~Ossher, and C.~Lopes.
\newblock Sourcerer: An internet-scale software repository.
\newblock In \emph{Proc. SUITE}, pages 1--4, 2009.

\bibitem[Barbosa et~al.(2012)Barbosa, Garcia, and Mezini]{heuristics}
E.A. Barbosa, A.~Garcia, and M.~Mezini.
\newblock Heuristic strategies for recommendation of exception handling code.
\newblock In \emph{Proc. SBES}, pages 171--180, 2012.

\bibitem[Buse and Weimer(2010)]{readability}
R.P.L. Buse and W.R. Weimer.
\newblock Learning a metric for code readability.
\newblock \emph{Softw. Eng., IEEE Trans.}, 36\penalty0 (4):\penalty0 546--558,
  2010.

\bibitem[Cabral and Marques(2007)]{fieldstudy}
B.~Cabral and P.~Marques.
\newblock Exception handling: A field study in java and .net.
\newblock In \emph{Proc. ECOOP}, pages 151--175, 2007.

\bibitem[Chang et~al.(2001)Chang, Jo, Yi, and Choe]{chang}
B.M. Chang, J.W. Jo, K.~Yi, and K.M. Choe.
\newblock Interprocedural exception analysis for java.
\newblock In \emph{Proc. SAC}, pages 620--625, 2001.

\bibitem[Garcia et~al.(2001)Garcia, Rubira, Romanovsky, and Xu]{garcia}
A.F. Garcia, C.M.F. Rubira, A.~Romanovsky, and J.~Xu.
\newblock A comparative study of exception handling mechanisms for building
  dependable object-oriented software.
\newblock \emph{JSS}, 59\penalty0 (2):\penalty0 197 -- 222, 2001.

\bibitem[Goodenough(1975)]{goodenough}
J.B. Goodenough.
\newblock Exception handling: Issues and a proposed notation.
\newblock \emph{Commun. ACM}, 18\penalty0 (12):\penalty0 683--696, 1975.

\bibitem[Holmes and Murphy(2005)]{strathcona}
R.~Holmes and G.C. Murphy.
\newblock Using structural context to recommend source code examples.
\newblock In \emph{Proc. ICSE}, pages 117--125, 2005.

\bibitem[Nguyen et~al.(2009)Nguyen, Nguyen, Pham, Al-Kofahi, and Nguyen]{groum}
T.T. Nguyen, H.A. Nguyen, N.H. Pham, J.M. Al-Kofahi, and T.N. Nguyen.
\newblock Graph-based mining of multiple object usage patterns.
\newblock In \emph{Proc. ESEC/FSE}, pages 383--392, 2009.

\bibitem[Posnett et~al.(2011)Posnett, Hindle, and Devanbu]{simpler}
Daryl Posnett, Abram Hindle, and Premkumar Devanbu.
\newblock A simpler model of software readability.
\newblock In \emph{Proc. MSR}, pages 73--82, 2011.
\newblock ISBN 978-1-4503-0574-7.

\bibitem[Rahman et~al.(2014)Rahman, Yeasmin, and Roy]{surfclipse}
M.M Rahman, S.~Yeasmin, and C.~Roy.
\newblock Towards a context-aware ide-based meta search engine for
  recommendation about programming errors and exceptions.
\newblock In \emph{SEW}, pages 194--203, Feb 2014.

\bibitem[Reimer and Srinivasan(2003)]{reimer}
D.~Reimer and H.~Srinivasan.
\newblock Analyzing exception usage in large java applications.
\newblock In \emph{Proc. EHOOS}, pages 10--19, 2003.

\bibitem[Robillard and Murphy(2003)]{robi}
M.~P. Robillard and G.C. Murphy.
\newblock Static analysis to support the evolution of exception structure in
  object-oriented systems.
\newblock \emph{TOSEM}, 12\penalty0 (2):\penalty0 191--221, 2003.

\bibitem[Roy and Cordy(2008)]{croy}
C.K. Roy and J.R. Cordy.
\newblock {NICAD}: {A}ccurate {D}etection of {N}ear-{M}iss {I}ntentional
  {C}lones {U}sing {F}lexible {P}retty-{P}rinting and {C}ode {N}ormalization.
\newblock In \emph{ICPC}, pages 172--181, 2008.

\bibitem[Shah et~al.(2008)Shah, G\"{o}rg, and Harrold]{shah}
H.~Shah, C.~G\"{o}rg, and M.J. Harrold.
\newblock Visualization of exception handling constructs to support program
  understanding.
\newblock In \emph{Proc. SoftVis}, pages 19--28, 2008.

\bibitem[Sinha et~al.(2004)Sinha, Orso, and Harrold]{sinha}
S.~Sinha, A.~Orso, and M.J. Harrold.
\newblock Automated support for development, maintenance, and testing in the
  presence of implicit control flow.
\newblock In \emph{Proc. ICSE}, pages 336--345, 2004.

\bibitem[Takuya and Masuhara(2011)]{selene}
W.~Takuya and H.~Masuhara.
\newblock A spontaneous code recommendation tool based on associative search.
\newblock In \emph{Proc. SUITE}, pages 17--20, 2011.

\end{thebibliography}

\end{document}